# Autonomous Code Review: Hidden-Test Results on GitHub Issues

**Adam Shakhabov**
theadamo86@gmail.com

## Abstract

Can an autonomous cross-provider review step change whether code written for a real issue passes hidden tests? We report a retrospective benchmark of 12 historical GitHub issues (five C#, five JavaScript, and two Python) from five repositories. Two directed author–reviewer pairs, OpenAI GPT-6.1-Sol and Anthropic Claude Sonnet 5.5, each ran twice per issue, yielding 48 planned workflows. A bounded terminal agent produced an initial patch; a cross-provider model reviewed it; the author model adjudicated findings and could revise the patch. The same hidden evaluator assessed initial and final patches. Forty workflows completed, while eight terminated before a complete evaluator outcome. Four completed workflows passed, and all four already passed before review. No initial-to-final pass/fail transitions occurred among completed workflows.

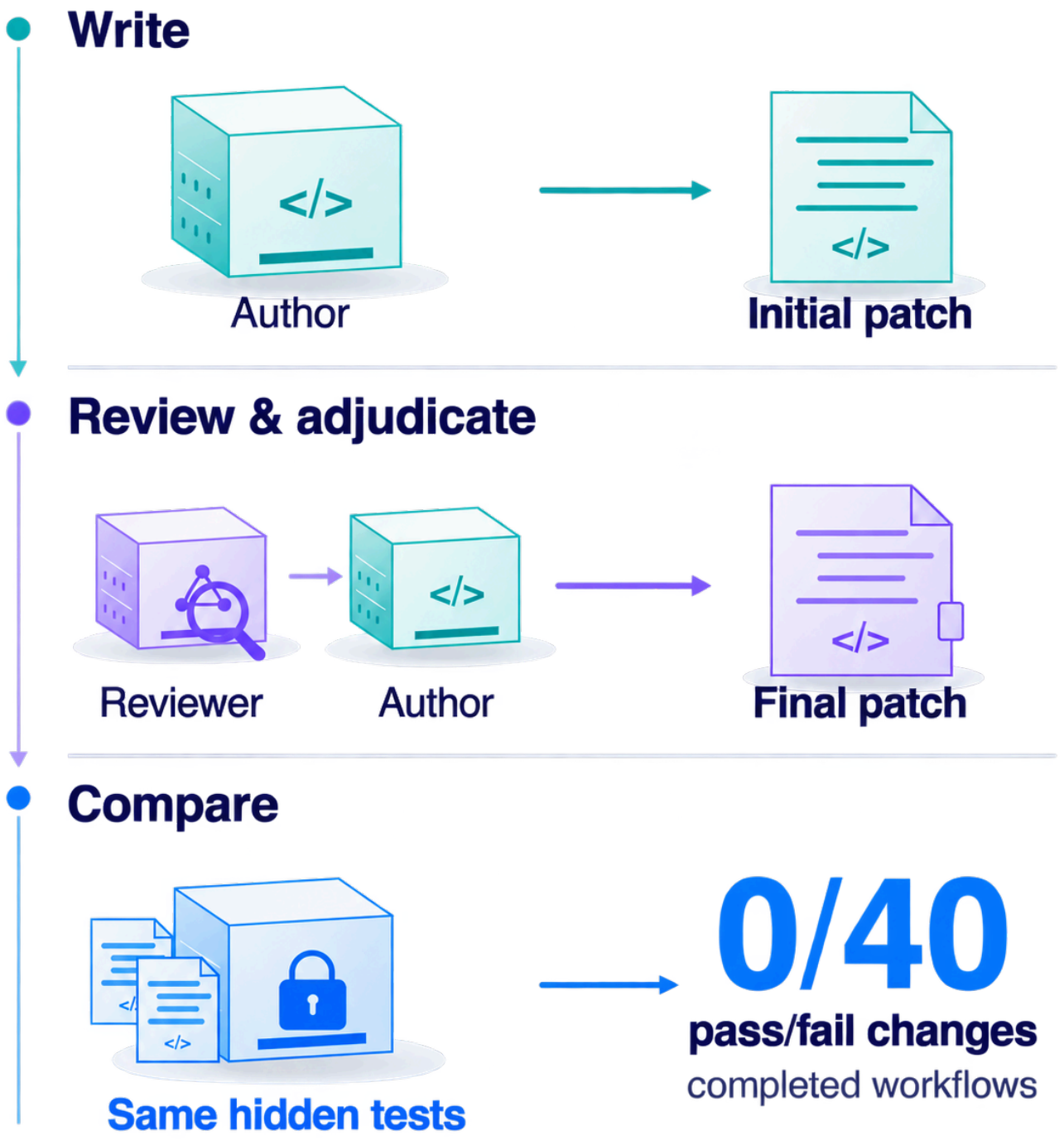

Figure 1: Paired evaluation: the initial and final patches are assessed with the same hidden-test oracle.

## 1 Introduction

Pull requests (PRs) combine a proposed code change with a pre-merge review process. The claim that PR review is obsolete in an LLM-assisted workflow is broader than a benchmark can establish: it concerns engineering time, defect prevention, coordination, and production outcomes. A retrospective agent benchmark can answer a narrower question. Given a historical issue and a pre-change repository state, does inserting autonomous cross-provider review and author-model adjudication change whether a test-based evaluator passes?

This paper reports a small retrospective benchmark. The primary comparison is within workflow: the author model's initial patch is evaluated, then reviewed and adjudicated, and the resulting final patch is evaluated again. This design observes whether the review/remediation bundle changes a binary hidden-test outcome for the same run. It does not compare against a randomized human-review arm, nor does it measure the total value or cost of the PR process.

The central result is deliberately narrow: **In this 12-issue, prior-exposed retrospective benchmark, autonomous cross-provider review and adjudication produced no observed initial-to-final changes in hidden-test pass/fail status across completed workflows.** Forty of 48 planned workflows completed. Of those 40, four passed and 36 failed; each pass was already present in the initial patch. The other eight workflows are reported as incomplete, not silently removed from the study flow.

Our contribution is a content-reduced public account of this workflow and its paired outcomes. The accompanying repository supports offline reproduction of numerical summaries from released outcome labels, but it does not provide raw trajectories or hidden evaluators for independent end-to-end replay or oracle validation. We do not claim that LLM review is useless, that the initial and final patches are statistically equivalent, or that PR review should be removed.

## 2 Related work

Repository-level benchmarks evaluate whether an agent can turn a task description and a repository state into a patch that satisfies tests. SWE-bench established a widely used task format built from real GitHub issues and corresponding pull requests, with repository-level execution tests as the outcome rather than patch-string similarity (Jimenez et al., 2024). SWE-agent subsequently showed that the agent-computer interface and interaction loop can materially affect repository-task performance, emphasizing that a benchmark measures a model together with its tools and scaffold, not the model in isolation (Yang et al., 2024). This is directly relevant to the evaluated workflow, whose bounded terminal interface is part of the treatment being measured.

Benchmark validity is a separate concern from agent capability. In SWE-bench Verified, software developers screened candidate instances for underspecified problem statements and potentially unfair tests; the authors report filtering out 68.3% of the annotated sample for one or more quality concerns (OpenAI, 2024). This study uses frozen historical tasks and test assets, but it did not conduct a comparable independent expert review of every issue–evaluator pair. We therefore treat hidden-test pass as an operational outcome, not a complete measure of issue resolution.

Work on LLM code review often measures whether a generated review identifies known issues. SWR-Bench, for example, evaluates review findings against annotated change points in real pull requests and reports precision/recall/F1-style measures (Zeng et al., 2025). Such detection metrics address whether a reviewer can produce useful findings. This study instead asks a downstream question: after an author model adjudicates those findings, does the binary test outcome change? The measures are complementary and should not be conflated.

Terminal-Bench likewise treats realistic command-line interaction as an agent-level capability and uses isolated task environments with task-specific verification (Merrill et al., 2026). Across these benchmark families, the environment, task specification, interaction policy, and verifier are all part of the measurement. The result should accordingly be read as a property of a specified simulated workflow—not as an intrinsic property of cross-provider review in general.

## 3 Research questions and estimands

We address two descriptive questions:

1. **RQ1 (resolution):** What proportion of the 48 planned workflows passed the hidden evaluator, and what proportion passed among workflows with a complete evaluator outcome?
2. **RQ2 (review-stage change):** Among completed workflows, how often did the hidden-test pass/fail status change between the initial author patch and the final patch after cross-provider review and adjudication?

For RQ2, the primary observed contrast is paired within a workflow. Let $Y_0$ be the evaluator outcome for the initial patch and $Y_1$ the outcome for the final patch, where pass is 1 and fail is 0. The observed transition rate is the count of workflows where $Y_1 n = Y_0$, divided by completed workflows. The net percentage-point uplift is the completed-workflow final pass rate minus its initial pass rate. These descriptive statistics do not identify the causal effect of human review or PR policy.

## 4 Study design

### 4.1 Corpus and unit of analysis

The frozen corpus contains 12 real GitHub issue contexts, each linked to a historical resolving pull request and a repository snapshot. The tasks span five repositories: ABP Framework, Kiota, Directus, Strapi, and Wagtail. The language allocation is five C# issues, five JavaScript issues, and two Python issues. The task selection was purposive and feasibility-driven, not a probability sample of open-source issues.

For each task, the frozen manifest records the issue/PR identifiers, base commit, source-tree hash, evaluator command, evaluator asset hashes, and dependency hashes. The repository snapshot is denoted C0. Historical resolution artifacts informed the test-based evaluation contract, but were withheld from the coding agents. Hidden evaluator assets were physically separated from the agent workspace. The execution environment used Bubblewrap isolation and disabled network access during agent and evaluator execution; dependencies were prepared locally.

The primary unit for generalization is the issue, not an individual model call. Each issue was run in two directed author–reviewer configurations, with two replicates per configuration. Thus, 48 workflows represent 12 issue clusters with repeated runs, not 48 independent software tasks. The exact task list appears in Appendix A.

### 4.2 Prior exposure and task inputs

The tasks had already been presented to the selected models in earlier pilot runs. This prior exposure can increase familiarity and possible training-data contamination risk. It also means this benchmark is not an independent replication of the corpus. We do not infer that a specific output was contaminated, and we do not use patch similarity as evidence of leakage.

To avoid giving models information added after a historical merge, task text followed the frozen input policy: where the current issue body had been updated after the resolving PR, the model received only the title; otherwise it received a bounded current issue snapshot. A prior corpus audit found title-only inputs in 9 of 12 tasks. This conservatism reduces future-information leakage but can make the task materially underspecified. We therefore report task-input policy as a validity threat rather than interpreting every failure as a model capability failure.

### 4.3 Models and workflow

The two model IDs were `openai/gpt-6.1-sol` and `anthropic/claude-sonnet-5.5`; the frozen configuration also records dated canonical slugs. Provider direction was crossed: OpenAI author → Anthropic reviewer and Anthropic author → OpenAI reviewer. Each direction covered all 12 tasks and two replicates.

The review/remediation bundle consisted of three sequential stages. First, the author model inspected the issue and repository and could read, search, edit, and run bounded local commands. The author-stage cap allowed up to 12 tool-enabled model turns and a separate final response, with bounded command counts, tool output, and request size. Second, a read-only cross-provider reviewer received a bounded rendering of the initial diff and returned a structured review; it had one model turn and no tools. Third, the author model adjudicated the findings and could revise the patch; the adjudicator had at most two turns and bounded tool access. Reviewer and adjudicator responses used strict JSON Schema structured outputs, with a maximum of ten findings enforced by schema validation and no automatic truncation. Completion-token limits were 400 for author requests and 800 for reviewer/adjudicator requests. Request-context compaction removed oldest complete tool exchanges when necessary while preserving the system prompt and original issue; the full tool trajectory remained in audit logs.

The initial and final patches were evaluated separately with the same task-specific hidden test command. Agents never received the hidden evaluator assets or results during authoring, review, or adjudication. The intervention assessed here is the entire post-author review/remediation bundle; the study does not separately randomize reviewer presence from adjudicator behavior.

### 4.4 Outcomes and analysis

A workflow passes only if the manifest-specified evaluator exits successfully. A failing test/build or an execution error is a fail when a complete evaluator result is present. A workflow that terminates before a complete final evaluator outcome is classified as incomplete, not converted into an observed evaluator fail for the completed-case transition analysis.

We report: (i) final pass count over all 48 planned workflows, with incomplete workflows counted as not-pass for this operational rate; (ii) pass count among completed workflows; (iii) the paired initial-to-final transition table among completed workflows; (iv) results by directed model pair; and (v) descriptive outcomes by task-input form. Initial outcomes for incomplete workflows are not treated as failures: in five reviewer/adjudicator terminations the initial evaluator ran but its result was not preserved, while three author-stage terminations occurred before initial evaluation. The full-planned initial pass rate is therefore unavailable. The frozen analysis code used an issue-cluster bootstrap with 2,000 resamples and seed 20261022. We retain its degenerate zero-uplift interval only as an algorithmic diagnostic; it is not interpreted as a confidence interval for the population effect.

We also report completion failures and API accounting. Unknown transport outcomes and unresolved reservations are not treated as known zero-cost calls. No workflow was retried after an unknown or terminal outcome.

## 5 Results

### 5.1 Completion and pass outcomes

All 48 planned workflow keys have unique durable records. Forty completed with a final evaluator outcome; eight terminated before a complete outcome. Among completed workflows, four passed and 36 failed. All four passes remained passes after review and adjudication; none of the 36 initial failures became a pass, and no initial pass became a fail. The paired transition table is shown in Table 1.

| Initial outcome | Final pass | Final fail | Total |
|---|---|---|---|
| Initial pass | 4 | 0 | 4 |
| Initial fail | 0 | 36 | 36 |
| **Total completed** | 4 | 36 | 40 |

Table 1: Initial-to-final hidden-test outcomes among the 40 completed workflows.

The final pass rate was 4/48 (8.3%) over the planned matrix, counting incomplete workflows as not-pass for this operational outcome; among completed workflows it was 4/40 (10.0%). The issue-weighted planned-matrix pass rate was also 8.3%, with a 95% issue-cluster bootstrap interval of 0.0%–20.8%. Two of 12 issues had at least one passing workflow (16.7%); no interval is reported for this distinct incidence measure. Among the 40 completed workflows, initial and final pass rates were both 4/40, yielding an observed uplift of 0 percentage points. The frozen bootstrap procedure mechanically returns [0.0, 0.0] for uplift because no issue had an observed transition; this degenerate output is retained only as a diagnostic and is not an inferential interval for effects on new issues. The full-planned initial pass rate cannot be recovered from the saved records.

### 5.2 Outcomes by task-input form

All four final passes arose in workflows for the three tasks supplied with a bounded current issue snapshot: 4/12 planned workflows passed (33.3%), and 4/10 completed workflows passed (40.0%). The nine title-only tasks had 0/36 planned passes and 0/30 completed passes. This is a descriptive stratification, not evidence that adding issue text caused success: input form is confounded with task identity, and the groups differ in their incomplete-workflow counts.

| Form | Planned | Complete | Pass / planned | Pass / complete |
|---|---|---|---|---|
| Title only | 36 | 30 | 0/36 | 0/30 |
| Bounded snapshot | 12 | 10 | 4/12 | 4/10 |
| **Total** | 48 | 40 | 4/48 | 4/40 |

Table 2: Final hidden-test pass outcomes stratified by the task-input form used in the frozen corpus.

### 5.3 Directed model pairs

| Pair | Complete | Initial pass complete | Final pass complete | Final pass planned |
|---|---|---|---|---|
| OA→AN | 18 | 0/18 | 0/18 | 0/24 |
| AN→OA | 22 | 4/22 | 4/22 | 4/24 |
| **Total** | 40 | 4/40 | 4/40 | 4/48 |

Table 3: Completion and pass outcomes by directed provider pair. OA denotes OpenAI; AN denotes Anthropic.

All four passes came from Anthropic-author → OpenAI-reviewer workflows: both replicates of `abp-multitenancy-26222` and both replicates of `replacement-strapi-transfer-27564`. In each, the initial patch already passed. No OpenAI-author workflow passed. These pair-level counts are descriptive and do not support a model ranking: the task sample is small, the runs are clustered by issue, and termination rates differed by direction.

### 5.4 Incomplete workflows and operational cost

Eight of 48 workflows lacked a complete evaluator outcome. Five terminated during reviewer/adjudicator structured-result validation, after the initial hidden evaluator had run; however, the stage-error handler replaced the record with a generic termination row, and the initial evaluator outputs were not persisted in the run directory. Their initial pass/fail outcomes are therefore unavailable in the saved data. Three other terminations occurred in the author stage before initial evaluation: two requests had unknown transport outcomes and one workflow was interrupted by the execution runner. The five missing initial results must not be recoded as failures. Terminations were asymmetric: six in OpenAI-author → Anthropic-reviewer and two in Anthropic-author → OpenAI-reviewer. This makes a completed-only analysis potentially selective.

The reported matrix recorded 564 API event records: 562 settled responses and two unknown transport events. Its settled API increment was \$4.072381; cumulative lifetime spend carried for-ward from earlier work was $10.181374 against the $50 ceiling. Open reservations at close were $0.1765 and are not counted as settled spend. These costs describe API usage only; they do not include engineering or research labor.

The bounded-history implementation compacted 141 of 396 recorded author turns, dropping 559 old tool exchanges from model-visible history while retaining full audit logs. Context compaction is part of the evaluated agent treatment and may affect its performance; it is not a neutral implementation detail.

## 6 Discussion

The primary finding is an absence of observed binary outcome change among the 40 completed workflows: four remained pass and 36 remained fail. Under the tested protocol, autonomous review and author-model adjudication did not rescue a hidden-test failure or undo an initial pass in those completed cases. This is not proof that the true effect is exactly zero. The mechanically degenerate bootstrap output reflects no observed within-issue transitions and is not an informative uncertainty bound for unseen tasks.

The distinction between review quality and downstream code change matters. A reviewer may produce a correct, useful finding that the author rejects, or a finding may point to a behavior not captured by the benchmark tests. Conversely, an accepted edit might improve maintainability without changing the binary evaluator outcome. The primary metric detects only pass/fail changes. It does not adjudicate review comments for correctness, measure patch quality beyond the chosen tests, or value latency, security, readability, or maintainability. The result should not be restated as "the reviewer found nothing".

A second implication concerns the author scaffold. A terminal agent can inspect and edit a repository, but its measured success depends on the task text, turn and token budgets, available commands, history policy, dependency setup, and stopping rules. SWE-agent's analysis of agent-computer interfaces and Terminal-Bench's task-specific environments reinforce that agent performance is an interaction between model and scaffold (Yang et al., 2024; Merrill et al., 2026). The title-only inputs for many tasks and the compacting history policy may have constrained task resolution. Because the evaluated workflow was designed after earlier pilots and reused their task contexts, the study cannot identify the independent causal effect of those scaffold changes.

The result also does not compare a post-merge autonomous workflow against a conventional human pre-merge PR. There was no human-review arm, randomized assignment, direct measure of engineering time, or production-quality outcome. Therefore this study cannot support the broad claim that PRs are obsolete.

## 7 Limitations

**Small, purposively selected corpus.** Twelve issues across five repositories are insufficient for broad generalization. Repeated runs improve description of stochastic variation on these tasks but do not create additional independent issues. The two Python tasks in particular cannot support language-specific claims.

**Prior exposure and contamination.** All issue contexts had appeared in earlier pilot runs. The same models were used across configurations. Prior exposure weakens independence and raises contamination risk. The study does not establish that leakage occurred, but it cannot exclude model familiarity with public issue/PR artifacts.

**Task specification and oracle validity.** Nine of 12 inputs were title-only; three received a bounded current issue snapshot. All four observed passes came from the latter group (4/12 planned workflows, versus 0/36 for title-only inputs), but this post hoc descriptive split is confounded by task identity and cannot establish a causal effect of prompt detail. Existing corpus-preparation records document C0-fail/historical-fix-pass checks for selected test contracts, summarized in Appendix B. These checks show sensitivity to the historical change, but not that the prompt communicated all requirements, that the historical fix is the only valid solution, or that each evaluator fully represents the issue. We did not independently assess task–test alignment through expert annotation comparable to SWE-bench Verified (OpenAI, 2024). A pass therefore means "passed this frozen evaluator," not "fully resolved every user need."

**Incomplete outcomes.** Eight workflows (16.7%) terminated before a complete outcome, with different rates by direction. Initial evaluator outcomes were run but not persisted for five reviewer/adjudicator terminations, and were not run for three author-stage terminations. The full-planned initial pass rate and full-planned initial-to-final uplift are not recoverable from saved records. Counting incomplete runs as non-passes is used only for the operational final pass rate; completed-only transitions condition on completion. Neither approach eliminates potential missingness bias.

**Bundled intervention and binary metric.** Reviewer and adjudicator were not independently manipulated. The study measures their combined post-author effect on one binary outcome. It does not estimate reviewer precision, recall, comment quality, human effort, or total engineering value. An evaluator can also miss correct alternative implementations or accept incomplete ones.

**Implementation-specific workflow.** The bounded terminal agent, API structured outputs, context compaction, and runtime recovery policy define the evaluated workflow. Results are not directly comparable across earlier pilot implementations because their protocols differed and their task contexts had already been exposed.

## 8 Conclusion

In this 12-issue, prior-exposed retrospective benchmark, autonomous cross-provider review and adjudication produced no observed initial-to-final changes in hidden-test pass/fail status across completed workflows. Four of 40 completed workflows passed, and all four had already passed before the review/remediation bundle. Eight of 48 planned workflows were incomplete and are reported separately. The outcome supports a narrow descriptive statement about one simulated system and one fixed corpus. It does not demonstrate equivalence, establish that LLM review is useless, compare autonomous review with human PR review, or show that "PR is dead."

A stronger follow-up would require a new, non-reused task set; independent expert validation that explicitly traces each evaluator requirement to information visible in the task prompt or repository; a prespecified missing-outcome policy; and measures beyond binary tests. Future runs should persist initial evaluator summaries before invoking review stages. To answer the broader PR-policy question, a prospective comparison with a human-review arm and direct measurements of time, defects, and remediation would be needed.

## References

Carlos E. Jimenez, John Yang, Alexander Wettig, Shunyu Yao, Kexin Pei, Ofir Press, and Karthik Narasimhan. 2024. SWE-bench: Can Language Models Resolve Real-World GitHub Issues?. *International Conference on Learning Representations*. https://arxiv.org/abs/2310.06770

Mike A. Merrill, Alexander G. Shaw, Nicholas Carlini, Boxuan Li, Harsh Raj, and others. 2026. Terminal-Bench: Benchmarking Agents on Hard, Realistic Tasks in Command Line Interfaces. *arXiv preprint arXiv:2601.11868*. https://arxiv.org/abs/2601.11868

OpenAI. 2024. Introducing SWE-bench Verified. Official research-project publication. https://openai.com/index/introducing-swe-bench-verified/

John Yang, Carlos E. Jimenez, Alexander Wettig, Kilian Lieret, Shunyu Yao, Karthik Narasimhan, and Ofir Press. 2024. SWE-agent: Agent-Computer Interfaces Enable Automated Software Engineering. *arXiv preprint arXiv:2405.15793*. https://arxiv.org/abs/2405.15793

Zhengran Zeng, Ruikai Shi, Keke Han, Yixin Li, Kaicheng Sun, Yidong Wang, Zhuohao Yu, Rui Xie, Wei Ye, and Shikun Zhang. 2025. Benchmarking and Studying the LLM-based Code Review. *arXiv preprint arXiv:2509.01494*. https://arxiv.org/abs/2509.01494

## A Frozen task set

Table 4 lists the 12 task identifiers, repository, language, and linked issue. Issue IDs are those recorded in the frozen corpus manifest; aliases are retained as used in the run artifacts.

| Run alias | Repository | Language | Issue | PR |
|---|---|---|---|---|
| `abp-multitenancy-26222` | `abpframework/abp` | C# | 26222 | 26223 |
| `replacement-abp-duplicates-26208` | `abpframework/abp` | C# | 26208 | 26209 |
| `kiota-8134` | `microsoft/kiota` | C# | 8133 | 8134 |
| `kiota-8100` | `microsoft/kiota` | C# | 8099 | 8100 |
| `kiota-8096` | `microsoft/kiota` | C# | 8095 | 8096 |
| `directus-locale-28141` | `directus/directus` | JavaScript | 28141 | 28147 |
| `directus-tooltip-28165` | `directus/directus` | JavaScript | 28165 | 28263 |
| `strapi-events-28016` | `strapi/strapi` | JavaScript | 28016 | 28017 |
| `replacement-directus-kv-28171` | `directus/directus` | JavaScript | 28171 | 28172 |
| `replacement-strapi-transfer-27564` | `strapi/strapi` | JavaScript | 27564 | 27565 |
| `wagtail-uuid-14462` | `wagtail/wagtail` | Python | 14462 | 14463 |
| `wagtail-sitemap-14704` | `wagtail/wagtail` | Python | 14704 | 14740 |

Table 4: Task identifiers and issue provenance in the frozen corpus manifest.

## B Task inputs and archived oracle checks

Table 5 reports the input form and the already archived pre-experiment C0/historical-fix validation evidence. These are records from corpus preparation, not new evaluator executions for this manuscript revision. "C0 fail / historical fix pass" indicates the selected hidden test failed on the base snapshot and passed after applying the historical production fix. The checks assess test sensitivity to that historical change; they do not establish prompt sufficiency or semantic completeness of the oracle.

| Task | Input | Archived pre-run check / caveat |
|---|---|---|
| `abp-multitenancy-26222` | Bounded snapshot | C0 fail / historical fix pass; repeated. Restore assets were prepared outside the isolated test phase. |
| `replacement-abp-duplicates-26208` | Bounded snapshot | Two later warm-cache checks: C0 fail / historical fix pass; earlier attempts failed during dependency setup. Eligibility/I0 provenance was incomplete in the audit. |
| `kiota-8134` | Title only | Two recorded checks: build passed; C0 test failed / historical-fix test passed. |
| `kiota-8100` | Title only | Two recorded checks: build passed; C0 test failed / historical-fix test passed. |
| `kiota-8096` | Title only | Two recorded checks: build passed; C0 test failed / historical-fix test passed. |
| `directus-locale-28141` | Title only | One recorded C0-fail / historical-fix-pass check. |
| `directus-tooltip-28165` | Title only | One recorded C0-fail / historical-fix-pass check. |
| `strapi-events-28016` | Title only | One recorded C0-fail / historical-fix-pass check. |
| `replacement-directus-kv-28171` | Title only | Two recorded C0-fail / historical-fix-pass checks. |
| `replacement-strapi-transfer-27564` | Bounded snapshot | Two recorded C0-fail / historical-fix-pass checks. |
| `wagtail-uuid-14462` | Title only | Two recorded C0-fail / historical-fix-pass checks. |
| `wagtail-sitemap-14704` | Title only | Two recorded C0-fail / historical-fix-pass checks. |

Table 5: Task input form and previously archived hidden-test contract checks.

## C Reproducibility record

The evaluation matrix completed after a safe continuation. The run manifest records 48 planned and 48 unique durable workflow rows, 40 completed workflows, and eight terminations. The frozen corpus manifest SHA-256 is `1c6f6cc850ea69afb34d1b8fe2e6feff7da47a399a6a389613431764b37fd3f6`. The raw `results.json` SHA-256 is `0ebb7c6a0f1db97ccecbae6c0015aa013f6c0675f92c39628e1a3981dab847cd`; the original private run-analysis JSON SHA-256 is `87ea4bfffa674e9f7eff5e69a4bea575692aa2ecf2c28f7220e32cdcf13596a7`. The public `data/analysis_v8.json` is a separately versioned, post hoc corrected summary: it distinguishes five initial evaluator results that were not persisted from three evaluations that never ran. The complete run manifest lists hashes for 289 original artifacts. These hashes identify the local study artifacts; the public release contains content-reduced data rather than the raw run bundle.

**Data and code availability.** The public release is available at github.com/AdamShakhabov/autonomous-code-review-study. The repository README links to a ZIP bundle containing the content-reduced workflow and accounting data, task provenance, frozen summaries and configuration, harness/analysis code, protocol, and manuscript source. The included offline verifier recomputes the reported numerical tables from the released outcome labels without model API calls or hidden-test execution. Raw model transcripts, tool traces, patches, API payloads, source snapshots, and hidden evaluator assets are excluded; consequently the release supports numerical reanalysis conditional on the reported evaluator labels, not independent end-to-end replay or oracle validation. The release assigns MIT to study-authored code and CC BY 4.0 to study-authored documentation and data; upstream issue and pull-request materials remain subject to their original terms.

**Ethics statement.** The study used public repository issues and model-generated code. It did not recruit human participants or collect private user data. Public repository content remains subject to its original license; any redistribution of source snapshots requires a license review.